\documentclass[reprint,final]{revtex4-1}

\usepackage{amssymb} % The amssymb package provides various useful mathematical symbols
\usepackage{amsthm}% The amsthm package provides extended theorem environments
\usepackage{amsmath}
\usepackage{graphicx,subfigure}% Include figure files
\usepackage{dcolumn}% Align table columns on decimal point
\usepackage{bm}% bold math
\usepackage{cancel}
\usepackage{float}
\usepackage{epstopdf}
\usepackage{hyperref}
\usepackage{url}
\usepackage[svgnames] {xcolor}

\hypersetup{colorlinks=true, linkcolor=blue, urlcolor=blue, citecolor=blue}

\begin{document}

\title{Robust Geo-Neutrino Results}

\author{S.T. Dye}
\affiliation{Department of Physics and Astronomy, University of Hawaii, Honolulu, HI, 96822}
\affiliation{Department of Natural Sciences, Hawaii Pacific University, Kaneohe, HI, 96744}

\date{\today}% It is always \today, today,

\begin{abstract}
\vspace{1mm}
\noindent
Geo-neutrino observations probe the quantities and distributions of terrestrial heat-producing elements uranium and thorium. The quantities of these elements gauge global radiogenic power, offering insights into the origin and thermal history of the Earth. The distributions reveal the initial partitioning and subsequent transport of these trace elements between metallic core, silicate mantle, and crust types. Ongoing observations at underground sites in Japan and Italy record the energies but not the directions of geo-neutrinos from uranium and thorium. Without directions pointing back to source regions, disentangling the signals from various reservoirs requires resolution of differing rates or energy spectra at separate sites. Due to limited statistics and site contrast, however, the observations at Japan and Italy do not yet measure distinct rates or energy spectra. Further analyses of the observations that derive fluxes, determine a signal from the mantle, and assess the global radiogenic power of uranium and thorium, depend on geochemical assumptions and model predictions. This letter discusses opportunities for eliminating or minimizing these dependencies through observations at dissimilar sites, producing robust geo-neutrino results. 
\end{abstract}
\maketitle

\section{Introduction} 
\begin{figure}
\centering
\includegraphics[trim = 5mm 50mm 25mm 40mm, clip, scale = 0.4] {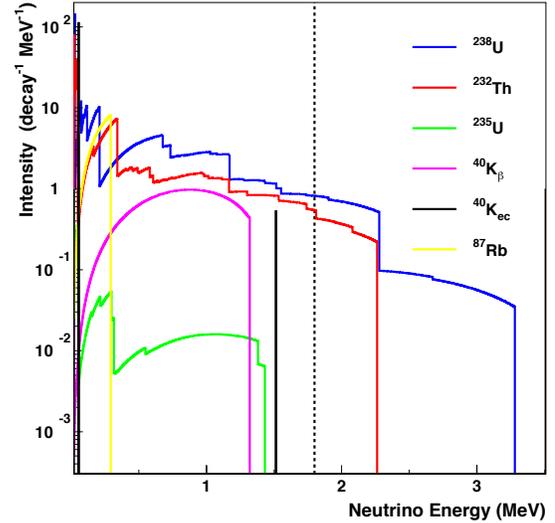}
\caption{Energy spectra of geo-neutrinos from $^{40}$K, $^{87}$Rb, $^{232}$Th, $^{235}$U, and $^{238}$U. The vertical dashed line marks the $1.8$ MeV threshold energy of \eqref{ibd}.}
\label{fig:nuspec}
\end{figure}

\begin{figure}
\centering
\includegraphics[trim = 5mm 50mm 25mm 40mm, clip, scale = 0.4] {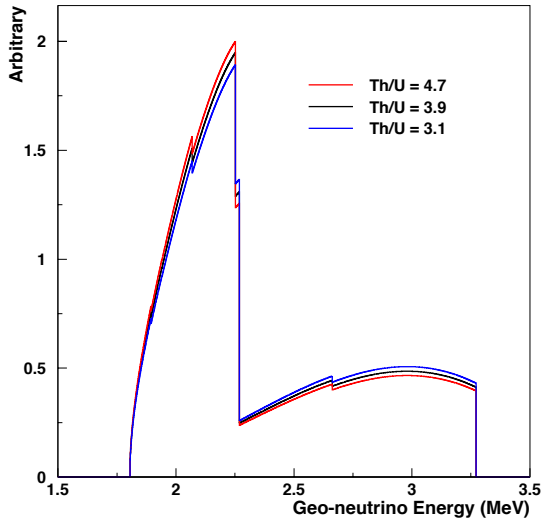}
\caption{Geo-neutrino energy spectrum detected by \eqref{ibd} for several values of Th/U. The area under each curve is normalized to one interaction.}
\label{fig:thurat}
\end{figure}

Electron antineutrinos from radioactive isotopes within the Earth are observed with high significance at two underground locations \cite{gando13,agostini15}. Radioactive isotopes with lifetimes comparable to the age of Earth and which initiate nuclear $\beta^-$ decays are candidate sources. These isotopes include $^{40}$K, $^{87}$Rb, $^{232}$Th, $^{235}$U, and $^{238}$U. Figure~\ref{fig:nuspec} shows the emission spectra of these isotopes. Multipurpose detectors at underground locations efficiently record the inverse beta decay of the free proton
\begin{equation}
\overline{\nu}_e + p \rightarrow n + e^+.
\label{ibd}
\end{equation}
\noindent
This quasi-elastic interaction requires antineutrino energy greater than 1.8 MeV. Consequently, these detectors are sensitive to $^{232}$Th and $^{238}$U but not to $^{40}$K, $^{87}$Rb, and $^{235}$U. The kinetic energy of the scattered positron in \eqref{ibd} is related to the antineutrino energy by $T_{e^+}=E_\nu-1.8$ MeV. This relationship allows measurement of the spectrum of the highest energy geo-neutrinos from $^{232}$Th and $^{238}$U. The shape of the measured energy spectrum depends on the relative contributions of $^{232}$Th and $^{238}$U, or equivalently the effective thorium to uranium elemental ratio (Th/U) of the geo-neutrino signal at the detection site. Figure~\ref{fig:thurat} shows energy spectra for several values of Th/U. The similar shapes of the spectra indicate the challenge of measuring different values of Th/U.

The measured rate of geo-neutrino interactions depends on the proximity of the observatory to the geochemical reservoirs of uranium and thorium. The initial inventory of uranium and thorium contents during formation of the Earth and any subsequent redistribution by various processes determine the mass fractions of these trace elements in the various reservoirs. Redistribution processes include geochemical fractionation between metallic core, silicate mantle, and crust types \cite{mcd95} and transport by geological and biological activity \cite{sleep2013}. Several other hypothetical processes \cite{boyet2005,boyet2006,campbell2012,wohlers15}, introduced to explain measurements of a greater samarium to neodymium elemental ratio of the silicate Earth compared with chondritic meteorites \cite{boyet2005}, could contribute but are less tenable in light of recent assessments {\cite{burk2016,bouvier2016}. The geochemical behaviors of uranium and thorium are not identical in the processes listed above. This affects the shape of the measured geo-neutrino energy spectrum at a given site.

The fundamental result from the geo-neutrino observations at Japan and Italy is the measured rate of interactions. It is simply the number of recorded interactions divided by the detector exposure, which is the number of target protons multiplied by the operation time. The measured rate of geo-neutrino interactions is free of geochemical assumptions and model predictions. A secondary result, which derives from the directly measured rate, estimates the fluxes of geo-neutrinos from uranium and thorium. Flux estimates depend on the shape of the two-component energy spectrum, which the observations are yet to resolve. To get past this, analyses find the best fit of the data to an assumed spectral shape. The assumed shape is given by Th/U~$=3.9$, which is the ratio found in chondritic meteorites. Fixing Th/U at 3.9 fixes the flux ratio $\phi$(Th)$/\phi$(U) at $0.85$. This geochemical assumption leads to potentially biased estimates of the observed fluxes. A tertiary result determines mantle fluxes using a prediction of site-dependent crustal fluxes and restricting geo-neutrino production to the silicate Earth, which excludes the metallic core. Mantle fluxes do not uniquely define mesospheric distributions of uranium and thorium. A further result, which depends on the mantle distributions, the predicted lithospheric distributions, and the assumed chondritic spectral shape, assesses global radiogenic power from uranium and thorium. The number of geochemical assumptions and predictions indicates the level at which geochemistry informs geo-neutrino results. This letter discusses opportunities for using geo-neutrino results to inform geochemistry.

\section{Observations}
Geo-neutrinos were observed initially at Japan \cite{kl05}. A total of $25^{+19}_{-18}$ geo-neutrino interactions were recorded during an efficiency-corrected exposure of ($4.87\pm0.24)\times10^{31}$ target proton years. A rate of $5.1^{+3.9}_{-3.6}\times10^{-31}$ per target proton per year was measured. Assuming the chondritic spectral shape (Th/U$=3.9$) established a $99\%$ confidence upper limit to the geo-neutrino flux from uranium plus thorium of $1.62\times10^7$ cm$^{-2}$s$^{-1}$. A subsequent observation of geo-neutrinos was made at Italy \cite{bellini10}, using an exposure of $1.52\times10^{31}$ target proton years ($252.6$ ton year $\times$ $6.02\times10^{28}$ target protons per ton). When adopting the same chondritic spectral shape, a total of $9.9^{+4.1}_{-3.4}$ geo-neutrino interactions were recovered, resulting in a measured rate of $6.5^{+2.7}_{-2.2}\times10^{-31}$ per target proton per year. Larger samples of geo-neutrino interactions, leading to more precise measurements, were recorded during continued observations at the sites in Japan and Italy \cite{kl11,bellini13,gando13,agostini15}.

The most recently reported estimates of the geo-neutrino flux due to uranium plus thorium are $3.4\pm0.8$ cm$^{-2}\mu$s$^{-1}$\cite{gando13} at Japan and $5.0\pm1.3$ cm$^{-2}\mu$s$^{-1}$\cite{agostini15} at Italy. Both estimates assume the chondritic ratio, giving $\phi$(Th)$/\phi$(U) $=0.85$. The magnitudes of these flux estimates are in good agreement with the predictions of a geochemical model \cite{huang13}. The model predicts $3.6^{+0.7}_{-0.5}$ cm$^{-2}\mu$s$^{-1}$ at Japan and $4.7^{+1.2}_{-0.9}$ cm$^{-2}\mu$s$^{-1}$ at Italy when assuming an average suppression of $<P_{ee}>=0.55$ due to neutrino oscillations. While both measurements assume a spectral shape given by Th/U~$=3.9$, the geochemical model predicts Th/U~$=4.0$ at Japan and Th/U~$=4.5$ at Italy. Figure~\ref{fig:ratthu} compares the observed and predicted uranium plus thorium fluxes and Th/U values of the fluxes at the observation sites.
\begin{figure}
\centering
\includegraphics[trim = 5mm 50mm 25mm 40mm, clip, scale = 0.4] {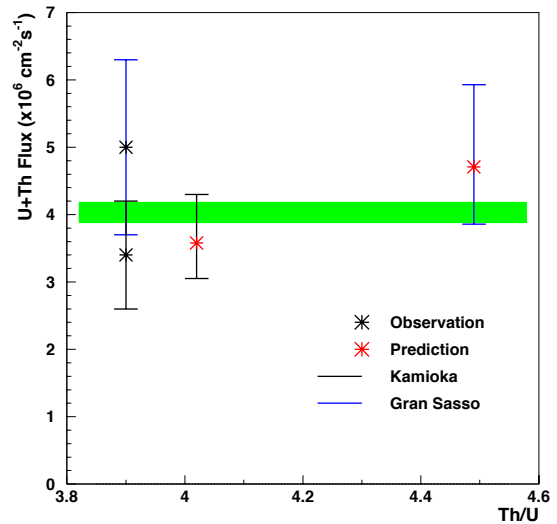}
\caption{Comparison of observations (black $\ast$) and predictions (red $\ast$) of the geo-neutrino flux due to uranium plus thorium versus Th/U for the observation sites at Kamioka, Japan (black line) and Gran Sasso, Italy (blue line). The green band shows the range of fluxes consistent with the observations and predictions.}
\label{fig:ratthu}
\end{figure}

\section{Independent Confirmation of Non-uniform Rates}
Analyses of geo-neutrino observations typically compare estimates of the uranium plus thorium fluxes with predictions of the non-uniform fluxes from crust types as determined by geochemical models. The differences are attributed to the fluxes from the mantle, excluding the core. For a potential mesospheric distribution of uranium and thorium, which is constrained by the predetermined mantle fluxes, analyses claim that geo-neutrino observations reveal a partial radiogenic heat model for Earth \cite{kl11,fior12}. An alternative interpretation of the observations, which is introduced as a potential geo-neutrino product, notes that the magnitudes of the estimated and predicted fluxes at the two sites are consistent within uncertainties (Fig.~\ref{fig:ratthu}). Consequently, the observations do not independently confirm the geochemical prediction of non-uniform uranium and thorium mass fractions. The null hypothesis becomes uniformly distributed mass fractions throughout spherical shells within the planet, producing a constant rate of geo-neutrino interactions from uranium and thorium. Presuming both observations are estimating the same surface flux suggests taking a weighted average of the two results. This calculation gives $\phi_{U+Th}=3.8\pm0.7$ cm$^{-2}\mu$s$^{-1}$ as the best estimate of the hypothetical constant surface flux.

Interestingly, the alternative interpretation of a constant surface flux is not ruled out by the surface heat flow measurement of $47\pm2$ TW \cite{davies10}. The preliminary reference earth model (PREM) \cite{dziew81} gives the density profile of a spherically symmetric Earth. According to the PREM, the geophysical response $G$ for the whole planet is $1.60\times10^9$ g cm$^{-2}$, while for the silicate mantle and crust it is $1.20\times10^9$ g cm$^{-2}$ \cite{dye12}. Using the isotopic activities of $^{238}$U and $^{232}$Th, $l_U=7.46\times10^4$ g$^{-1}$s$^{-1}$ and $l_{Th}=1.62\times10^4$ g$^{-1}$s$^{-1}$ \cite{dye12}, respectively, estimates the mass fraction of uranium, 
\begin{equation}
a_U=\frac{\phi_{U+Th}}{(l_U+l_{Th} \mathrm{Th/U}) <P_{ee}> G}.
\end{equation}
Corresponding mass fractions of thorium and potassium follow by scaling, according to $a_{Th}=a_U$Th/U and $a_K=a_U$K/U. The following example assumes Th/U$=3.9$ to be consistent with the geo-neutrino observations and K/U$=12000$. If the heat-producing elements are distributed uniformly throughout the whole planet, the mass fractions are $a_U=32\pm6$ ng/g, $a_{Th}=124\pm22$ ng/g, and $a_K=.038\pm.007$ wt\%. If the heat-producing elements are distributed uniformly throughout the silicate portion of the planet and absent from the metallic core, the mass fractions are $a_U=42\pm8$ ng/g, $a_{Th}=165\pm29$ ng/g, and $a_K=.051\pm.009$ wt\%. Using the PREM masses of the whole Earth ($M = 5.97\times 10^{27}$ g) and the silicate Earth ($M_{SE} = 4.03\times 10^{27}$ g} and applying the specific heat productions of uranium, thorium, and potassium, $h_U=9.85\times10^{-8}$ Wg$^{-1}$, $h_{Th}=2.63\times10^{-8}$ Wg$^{-1}$, and $h_K=3.33\times10^{-12}$ Wg$^{-1}$ \cite{dye12}, respectively, assesses the radiogenic power of the hypothetical distributions. A uniform distribution of heat-producing elements throughout only the silicate portion of the planet gives $41\pm7$ TW of radiogenic power. A uniform distribution throughout the whole planet gives $46\pm8$ TW of radiogenic power. Both power values are consistent with the measured surface heat flow of $47\pm2$ TW \cite{davies10} and, therefore, with a fully radiogenic heat model for Earth. When considering the whole planet distribution the core receives radiogenic heating per unit mass of $7.6\pm1.4$ pW kg$^{-1}$, which is consistent with estimates of the power requirements for the geodynamo \cite{pozzo}. 

Independent confirmation of non-uniform distributions of uranium and thorium may require observations at sites with greater difference between the predicted fluxes, or site contrast. A simple study estimates the exposures at selected sites that are needed to measure a geo-neutrino interaction rate different from the constant rate of the null hypothesis. The hypothetical constant flux of $3.8\pm0.7$ cm$^{-2}\mu$s$^{-1}$ corresponds to an interaction rate from \eqref{ibd} of $3.4\pm0.6\times10^{-31}$ per target proton per year when setting Th/U~$=3.9$. Geo-neutrino rates are sometimes expressed in terms of the terrestrial neutrino unit (TNU), where $1$ TNU equals $10^{-32}$ per target proton per year. An inverse TNU is a unit of exposure, where $1$ TNU$^{-1}$ equals $10^{32}$ target proton years. The best estimate of the rate, corresponding to the constant surface flux, is $34\pm6$ TNU.

\begin{figure}
\centering
\includegraphics[trim = 5mm 50mm 25mm 40mm, clip, scale = 0.4] {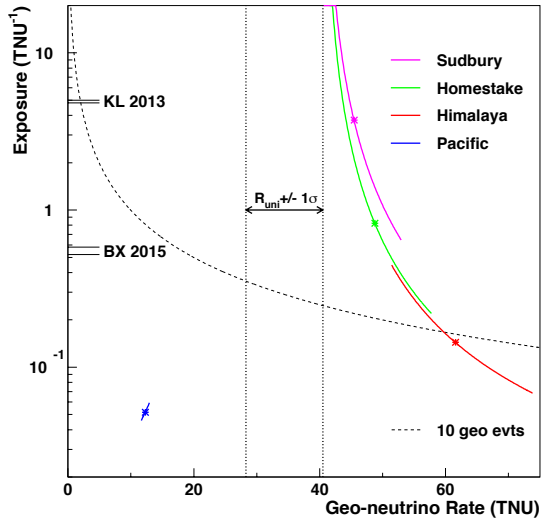}
\caption{Each colored curve estimates the exposure at a selected site that is required to independently confirm the geochemical prediction of non-uniform uranium and thorium mass fractions by measuring a geo-neutrino rate that is not consistent ($1\sigma$) with the best estimate of the uniform surface rate ($R_{uni}=34.4\pm6.1$ TNU). Vertical dotted lines mark the range of the measured rate. The horizontal ranges of the curves with the $\ast$ symbols represent the rates $\pm1\sigma$ predicted by a geochemical model \cite{huang13}. The black dashed curve follows the exposure needed for 10 geo-neutrino events. For reference, horizontal dotted lines next to the vertical axis mark the exposures of the most recently reported geo-neutrino observations at Japan (KL 2013) \cite{gando13} and Italy (BX 2015) \cite{agostini15}.}
\label{fig:rate_diff}
\end{figure}

Three continental sites and one oceanic site serve as examples in the study. The geo-neutrino rates, including uncertainties, at these prospective sites are predicted by a geochemical model \cite{huang13}. These predictions define the range of the geo-neutrino signal rate ($S$). The study assumes the background rate ($R$) results only from nuclear reactors. A recent appraisal of the site-specific reactor rates in the relevant geo-neutrino energy range assesses this background \cite{baldoncini}. A figure of merit given by $S/\sqrt{S+R}$ expresses the error on the geo-neutrino rate measured at the site. Figure~\ref{fig:rate_diff} shows the estimated exposures at each site that are needed to measure a geo-neutrino rate that is different from the hypothetical constant rate. As expected, the sites with the largest difference between predicted rates, or exhibiting the greatest contrast, require the smallest exposures. The required exposures at these sites ($\lesssim1$ TNU$^{-1}$) appear to be well within reach of future observations when compared with exposures already realized at existing sites.

\section{Independent Confirmation of Differential Mass Fractions}
The shape of the energy spectrum of geo-neutrinos initiating \eqref{ibd} is determined by the relative contributions of $^{232}$Th and $^{238}$U, or equivalently by the effective elemental ratio of thorium to uranium Th/U. According to a geochemical model \cite{huang13}, which predicts Th/U~$=4.0$ at Japan and Th/U~$=4.5$ at Italy, the energy spectrum varies from site to site because of differential uranium and thorium mass fractions. Due to limited statistics and site contrast, the existing geo-neutrino observations  at Japan \cite{gando13} and Italy \cite{agostini15} do not resolve the Th/U variation expected from the initial partitioning and subsequent redistribution of uranium and thorium between metallic core, silicate mantle, and crust types.

\begin{figure}
\centering
\includegraphics[trim = 5mm 50mm 25mm 40mm, clip, scale = 0.4] {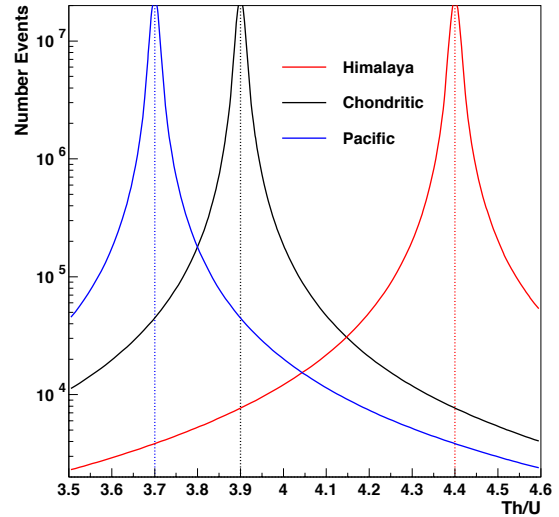}
\caption{Estimated number of events needed at each site to distinguish between different Th/U ratios according to \eqref{ntot}. The chondritic ratio of $3.9$ (vertical black dotted line) defines the spectral shape assumed by existing geo-neutrino flux measurements (black curves). At the selected continental site (Himalaya- red curves) the predicted ratio is $4.4$ (vertical red dotted line), while at the selected oceanic site (Pacific- blue curves) the predicted ratio is $3.7$ \cite{huang13}. The estimated number of events at both sites needed to resolve these selected values is found at the intersections of the curves and the vertical dotted lines.}
\label{fig:thu_him}
\end{figure}

Using contrasting sites, according to a geochemical model \cite{huang13}, a simple study estimates the number of events needed at each site to distinguish between different values of Th/U. A continental site and an oceanic site serve as examples. The effective thorium to uranium elemental ratios, which are predicted by the geochemical model \cite{huang13}, are Th/U~$=4.4$ at the continental site and Th/U~$=3.7$ at the oceanic site. Separating the normalized spectrum for a given value of Th/U at the endpoint energy of the $^{232}$Th geo-neutrinos ($E_{max}=2.254$ MeV), defines high and low energy regions. While the high energy region is due solely to uranium, the low energy region is due to thorium and uranium. As shown is Fig.~\ref{fig:thurat} the areas of these regions see-saw about the dividing energy as Th/U changes value. The normalizing condition requires $f_h+f_l=1$, where $f_h$ is the fractional area above the dividing energy and $f_l$ is the fractional area below the dividing energy. Using Poisson statistics to determine uncertainties simplifies the estimate of the  number of events needed at each site to distinguish between different values of Th/U. In terms of the fractional areas $f$ and $g$ and assuming perfect separation of the high and low energy regions, then resolution at $1\sigma$ is possible with a number of events
\begin{equation}
N=\Bigg[\frac{g_l/g_h\sqrt{1/g_l+1/g_h}+f_l/f_h\sqrt{1/f_l+1/f_h}}{f_l/f_h-g_l/g_h}\Bigg]^2.
\label{ntot}
\end{equation}
Resolution at $2\sigma$ ($3\sigma$) requires $4$ ($9$) times this number. Figure~\ref{fig:thu_him} shows the results of applying \eqref{ntot} to the Th/U values predicted at the contrasting sites. An estimated $4000$ events are required to resolve at $1\sigma$ different shapes of the energy spectra at the continental and oceanic sites. Assuming the geo-neutrino rates predicted at these sites, the corresponding exposures are $>60$ TNU$^{-1}$ at the continental site and $>300$ TNU$^{-1}$ at the oceanic site. These exposures are factors of $10$ and $50$, respectively, larger than any realized to date.

\section{Predicting Crustal Rates}
Conventional analyses of geo-neutrino observations determine mantle fluxes by subtracting predicted crustal fluxes from the estimated total fluxes. Both statistical and systematic uncertainties contribute to the error on the determined mantle fluxes. Systematic uncertainties come from the predicted crustal fluxes and a possible bias from assuming a spectral shape. Statistical uncertainties come from the estimated total fluxes. As the statistical precision of geo-neutrino observations improves with increased exposures, the systematic uncertainties begin to limit the precision of the determined mantle fluxes. Improved precision results from taking the quadratic sum of errors on presumably independent predictions of the fluxes from nearby and far away crust portions\cite{kl11,fior12}. While the same geochemical model suitably predicts the far away portion at observation sites, the nearby portions require individual predictions. Geochemical investigations provide these predictions at several observation sites \cite{colt11,huang14}. At other observation sites, predictions rely on the geochemical model, prompting recommendations for separate investigations at each site \cite{han16,sramek16}. 

An alternative method, which minimizes geochemical assumptions and predictions, predicts the crustal geo-neutrino interaction rates. It uses the geo-neutrino observations to inform geochemistry. This method requires the geochemical assumption that the rate of observed geo-neutrinos from the mantle is constant near the Earth's surface. A geochemical model, which predicts mantle fluxes at various sites that are within $\pm3\%$ \cite{huang13} of each other, supports this assumption. 

The observed rate of candidate geo-neutrino interactions is $T = C + M + R + B$, where $T$ is the total rate, $C$ is the crustal rate, $M$ is the mantle rate, $R$ is the rate from nuclear reactors, and $B$ is the rate from non-neutrino background sources. The latter two are typically estimated with good precision \cite{agostini15} and can be subtracted reliably to assess the geo-neutrino signal rate, $S = C + M$. Taking observations at two sites and imposing the assumption of a constant mantle rate gives the crustal rate at one site, $C_0$, in terms of the crustal rate at the other site, $C$, and the signal rates at the two sites, $C_0 = C + S_0 - S$. Assuming independent errors, the uncertainty of $C_0$ is 
\begin{equation}
\delta C_0 = (\delta C^2 + \delta S_0^2 + \delta S^2)^{1/2},
\label{error}
\end{equation}
where $\delta C$ is to be minimized and $\delta S_0$ and $\delta S$ are primarily determined by Poisson statistics. In the limit of infinite statistics $\delta C_0 \simeq \delta C$, showing that in principle the crustal rate at any site with a geo-neutrino observatory can be determined with precision approaching that of the site with the smallest uncertainty on the predicted crustal rate.

\section{Discussion}
The results of the presented studies guide estimates of the operation time required by various geo-neutrino detection projects to independently confirm the geochemical prediction of non-uniform and differential uranium and thorium mass fractions. Calculated operation times assume that each detector has a target density of $7.25\times10^{28}$ protons per tonne. Detection sites with target masses include projects on the continent: Sudbury with 780 tonnes \cite{chen06}, Himalaya with 3 kT \cite{sramek16}, and Homestake with 30 kT \cite{theia}; and a project in the ocean: Pacific with 10 kT \cite{dye06}. Assuming measured rates are those predicted by a geochemical model \cite{huang13}, the approximate operation times to exclude the null hypotheses of constant rate and constant Th/U are listed in Table \ref{tab:conrat}. These rough estimates result from simple studies, suggesting more sophisticated analyses are needed for planning observations that independently confirm the geochemical prediction of non-uniform and differential uranium and thorium mass fractions.

The Sudbury project is most likely to begin operation before projects at the other sites \cite{chen06}. However, the estimated operation time for this project to exclude the constant rate hypothesis approaches 10 years. Given the relatively short operation times estimated at other sites ($<1$ year), there is no clear prediction on which project is most likely to exclude the null hypothesis of a constant geo-neutrino rate. However, the prospects appear to be good for measuring a variation in the surface rate of geo-neutrinos within the next decade. The suggested measurement of this variation is free of geochemical assumptions and predictions. Such a geo-neutrino measurement would constitute an independent confirmation of non-uniform mass fractions of uranium and thorium.

Resolution of different geo-neutrino spectral shapes requires larger exposures than ruling out the constant geo-neutrino rate hypothesis, according to study herein. Exposures requirements at oceanic sites appear more demanding than at continental sites due to smaller predicted rates in the ocean. This study suggests that an oceanic detector of at least $30$ kT is needed to complete the measurement in $\sim10$ years. A detector of the same size deployed at a continental site completes the measurement in $\sim3$ years. Note, however, that resolution of different geo-neutrino spectral shapes requires measurements at both sites. Since the effective observation time is the longer of the two, the continental detector could be $\sim3$ times smaller than the oceanic detector without delaying the result. The suggested measurement of the variation in the spectral shapes is free of geochemical assumptions and predictions. Such a geo-neutrino rmeasurement would constitute an independent confirmation of differential mass fractions of uranium and thorium.
\begin{table}
\caption{Approximate operation times in years for detection projects at various sites to test the constant rate and constant Th/U hypotheses, assuming the rates predicted by a geochemical model \cite{huang13}. A minimum time required to collect at least 10 geo-neutrino events is imposed on the projects at Himalaya and Pacific for the test of the constant rate hypothesis. Resolution of different spectral shapes supposes exposures at a continental site and an oceanic site. Operation times are for Th/U~$=4.4$ at continental sites and Th/U~$=3.7$ at the oceanic site.}
\begin{tabular} {l c c c c}
\hline\noalign{\smallskip}
Hypothesis      & Sudbury & Himalaya  & Homestake & Pacific \\
\hline\noalign{\smallskip}
Constant rate  & 1-30 & 0.1-0.2 & 0.1-1 &  1 \\
Constant Th/U & 100 & 20 & 3 & 30 \\
\hline\noalign{\smallskip}
\end{tabular}
\label{tab:conrat}
\end{table}

A significant source of uncertainty associated with geo-neutrino results, particularly assessments of radiogenic heating, comes from the geochemical prediction of the crustal rates at observation sites. An alternative method of predicting the these rates uses geo-neutrino observations to reduce this uncertainty. Although this method requires a crustal rate prediction and the assumption of a constant mantle rate, it is otherwise free of geochemical assumptions and predictions. The following example using \eqref{error} outlines the potential reduction in uncertainty. According to a geochemical model, the uncertainty on the predicted uranium plus thorium crustal rates at observation sites near the middle of the Pacific Ocean is less than $1$ TNU \cite{huang13}. With reasonable statistics ($\delta S_0$ and $\delta S \lesssim 1$ TNU), an oceanic detector \cite{dye06} could estimate the mantle rate at a level of $\sim10\%$ and predict the crustal rates at continental observation sites \cite{chen06,sramek16} at a level of $\sim5\%$. Existing predictions of the crustal rates at continental sites achieve a precision of no better than $\sim15\%$ \cite{colt11,huang14}. The suggested alternative method supports the use of geo-neutrino observations to inform geochemistry. 

\section{Conclusions}
Existing geo-neutrino observations at Japan and Italy do not yet resolve different geo-neutrino rates or energy spectra. These observations are consistent with a geo-neutrino signal with a rate and spectral shape constant everywhere near the surface of the Earth. The best estimate of the constant rate is not excluded by measurements of the surface heat flow. An anticipated update of the observation at Japan is not expected to significantly change these results. This letter explores the exposure requirements for resolving different rates and spectral shapes with measurements at continental and oceanic detection sites. It finds that exposures of a few kT-y resolve different rates. The exposures to resolve different spectral shapes are about a few 100 kT-y. Tests of the constant surface flux and the constant spectral shape hypotheses are free of geochemical assumptions and model predictions. They provide independent confirmation of the non-uniform and differential mass fractions of uranium and thorium that are predicted by geochemistry. A suggestion to use geo-neutrino observations for predicting crustal rates provides an alternative to geochemical modeling. Observations at contrasting sites, such as the Himalayas and the middle of the Pacific Ocean, facilitate these robust geo-neutrino results.

\section*{Acknowledgments}
This research was inspired and supported by the Tohoku University Research Center for Neutrino Sciences and the Tohoku Forum for Creativity. Past support from the Cooperative Studies of the Earth's Deep Interior (CSEDI) program at the National Science Foundation (EAR 0855838 and EAR 1068097), the Cooperative Institute for Dynamic Earth Research (CIDER), and the Trustees' Scholarly Endeavors Program at Hawaii Pacific University contributed as well.

\newpage
\bibliographystyle{apsrev4-1}
\nocite{apsrev41control}
\bibliography{Nugeo_bib,revtex-custom}

\end{document}